\documentclass[submission,copyright,creativecommons]{eptcs}
\providecommand{\event}{F-IDE 2014} 
\usepackage{breakurl}             

\def\focalize{FoCaLiZe \mbox{}}

\usepackage{tikz}
\usetikzlibrary{arrows}
\tikzstyle{ent}=[circle,draw,thick,inner sep=0pt,minimum size=2.5mm]

\usepackage{graphicx}

\title{Teaching Formal Methods and Discrete Mathematics}

\author{Mathieu Jaume$^{1,2}$
\institute{1. Sorbonne Universit\'es, UPMC Univ. Paris 06, \\UMR 7606,
  LIP6, F-75005, Paris, France \\ 2. CNRS, UMR 7606, LIP6, F-75005, Paris, France}
\email{Mathieu.Jaume@lip6.fr}
\and Th\'eo Laurent
\institute{Sorbonne Universit\'es, UPMC Univ. Paris 06,\\ F-75005, Paris, France}
\email{Theo.Laurent@etu.upmc.fr}
}
\def\titlerunning{Teaching Formal Methods and Discrete Mathematics}
\def\authorrunning{M. Jaume, T. Laurent}

\begin{document}
\maketitle

\begin{abstract}
Despite significant advancements in the conception of (formal) integrated
development environments,
applying formal methods in software industry is still perceived as
a difficult task. To make the task easier, providing tools that help
during the development cycle is essential but we think that
education of computer scientists and software engineers is also an
important challenge to take up. Indeed, we believe that
formal methods courses 
do not appear sufficiently early in compter science curricula and thus
are not widely used and perceived as a valid professional skill.
In this
paper, we claim that teaching formal methods could be done at the
undergraduate level by mixing formal methods and discrete mathematics
courses and we illustrate such an approach with a small development within FoCaLiZe.
We also believe that this could considerably benefit the learning of
discrete mathematics. 
\end{abstract}

\section{Introduction}

Nowadays,
critical systems are evaluated according to some security standards like the
Common Criteria~\cite{CC2} or  according to safety ones like
the EN50128 for railways. To reach their  high-level rates,
these standards 
require the use of formal methods in order to ensure 
that security and safety requirements are effectively
satisfied by these systems.
Indeed, for large developments, ad hoc approaches have proven to be inadequate to ensure
that the delivered software truly satisfies safety and security
requirements. In fact,
the lack of formalisation often leads to produce systems whose
behaviors are not fully and precisely understood and described.
Formal methods aim at helping to build systems with high
safety and security assurances, and
formal integrated development environments (F-IDE) embed a variety of such
formal methods to help to specify, to document,
to implement, to test, to prove or to analyse critical systems.
Of course, such environments often ease
(and partially automate)
the application of formal methods during the development cycle, but
developing (and evaluating) critical systems is still a
difficult task that 
requires advanced technical knowledge and large amounts of time.
This is certainly one of the reasons why formal methods
are still not sufficiently used in industrial software
development. 

Developing F-IDE that ease the application of
formal methods is still  a challenging issue but
developping a F-IDE
which helps to learn formal methods is also a true challenge.
We believe that education is the corner stone
to promote the use of formal methods in
the software creation process.
The formal methods community has not enough focused its
attention to the education of computer scientists and software
engineers, especially at the undergraduate level.
Indeed, many computer science curricula
do not contain formal methods courses, or such 
material is not introduced sufficiently early. 

Presently, almost all these curricula include discrete mathematics courses but
often in isolation from computer science, leaving students
understanding little about why (and how) mathematics applies to computer science
and {\it vice versa}. Moreover,
teaching discrete mathematics is still often done in the traditional
way, using pen and paper, and
many computer science students are rather
``math-averse''
(they
are more familiar with ASCII characters than with greek alphabet!), perceive mathematics as a difficult discipline and
don't  understand its relevance in their curricula.

To address this issue, some discrete mathematics courses 
use functional programming languages (such as ML, OCAML, HASKELL, etc.)  to reinforce mathematical concepts.
There exists now some discrete mathematics
textbooks~\cite{Doets04,books/daglib/0007497,bookvandrunen} based on
such an approach whose
benefits are discussed 
in~\cite{DBLP:conf/sigcse/Wainwright92,oai:CiteSeerXPSU:10.1.1.19.3780,darosa02,Henderson02,DBLP:conf/oopsla/VanDrunen11,DBLP:conf/icfp/Page03}.
In~\cite{DBLP:conf/oopsla/VanDrunen11}, the author goes further 
by considering that computer science is also a vital topic for
contemporary mathematics students and that they will 
need some level of competency in programming at some point in their
professional practice. Hence, the author claims
that the
integrated work of mathematics and
computer science educators could considerably improve the learning of
both subjects: putting functional programming and
discrete mathematics in the same course provides a useful service for
both computer science and mathematics students. In fact,
functional programming languages are high-level languages
and thus are well suited to teach discrete
mathematics. Indeed, 
they permit to implement mathematical concepts without considering low level
issues such as data representation and memory allocation. Hence, 
mathematical notions can be easily introduced together with their 
implementations (that remain very
close to the concepts that get implemented) and can be manipulated
by students.
This is  a true way to reinforce
their understanding of mathematical concepts.
The benefit is also great  on the programming side. Using a
programming language to learn mathematical concepts leads to handle
these concepts as a specification for the program under development
and introduce students to the formal specification world. Then,
reasoning on the specification and the associated program is a way to smoothly
introduce the students to induction, logics and semantics, all notions
needed to demonstrate that a program meets its specification.
The goal on the computer science
side
is to put the emphasis on the correctness of the  computation
which is one of the main purposes of formal methods.
Currently, when they are used,  programming languages only serve
as a formalism to manipulate
the computational part of mathematical objects but not to express specifications or to
implement proofs. 
This may lead students to view formal methods as {\it a posteriori}  methods in the
programming tasks. In this paper, we claim that formal methods also provide 
an {\it a priori} help during the conception of software
that can be
taught in discrete mathematics courses: 
specifying a hierarchy of mathematical
discrete structures is a good introduction to the design of software architecture.

Even if proof assistants seem now to be mature enough to be adaped to the education,
at undergraduate level, formal reasoning is seldom introduced and
mostly appears in ``pure'' logic courses. For example, 
in~ \cite{hendriks-adn10}, the design of a web interface for Coq used to teach
logic to undergraduate students is presented. 
In the context of computer science teaching, formal reasoning is generally introduced at
a more advanced stage. This can be done by implementing some automated
theorem proving techniques (like in \cite{Harrison09}) or by using
proof assistants such as Coq or Isabelle.
In this case, F-IDE and theorem proving are not objects of the study but are
rather considered 
as a framework for teaching something else. Hopefully, using a language
as a vehicle for reinforcing concepts inevitably leads to learn some
methodological  and practical knowledges
about it.
For example, \cite{NKtoappear} is a semantics
textbook (to master students)  which is entirely based on the proof assistant
Isabelle.
The main benefit of using a proof assistant in
the teaching of semantics is that it allows students to experiment their
specifications and to make
proofs by using a
computer program, which guides them through the development of
a completely correct proof and gives them immediate feedback.
This avoids 
students to produce ``almost-but-not-completely-right proofs'' (as
called by Pierce in \cite{DBLP:conf/icfp/Pierce09}) or even worse ``LSD trip proofs''
(as called by Nipkow in~\cite{Nipkow-VMCAI12}).

As we said, we think that teaching formal methods to beginners
is essential to disseminate their use in the software industry. However, at
the undergraduate level, no prerequesites on computer science can be
assumed and we can only suppose some very basic knowledges in
mathematics that are also considered as prerequisites for the first
courses of discrete mathematics. Hence, we believe that using a F-IDE
could be helpful to teach both computer science and discrete
mathematics in a mixed course.

This paper aims at presenting our pedagogical approach of both
disciplines through a small mathematical development.
In this context, the F-IDE used as a teaching tool
must be suitable to express specifications
(i.e. properties), to write programs (i.e. definitions) and to make
proofs. One of the main issues is concerned with proofs. Within most theorem
provers, proofs are sequences of commands (belonging to a scripting
language)  that are hard to read for the
human: they lack the information what is
being proved at each point, and they lack structure.
Such provers are clearly not suitable to teach discrete mathematics at
the undergraduate level
since they do not provide a
proof language
close to the informal language of mathematics.
Furthermore, the proof language used must be abstract enough
to avoid to teach the fine structure of logic (the inference rules)
and to automate the ``trivial'' steps of proofs by allowing students
to only express what intermediate steps
might help the proof assistant to complete proofs.
For these reasons, we think that  the \focalize\cite{foc03} F-IDE is a good candidate
to teach both computer science and discrete mathematics at the
undergraduate level.
Indeed, \focalize is an object-oriented programming environment that
combines specifications, programs and proofs in the same language, and
permits declarative proof descriptions inspired by Lamport's
work~\cite{Lamport95,chaudhuri:proof}. 
These
features can be used to
formally express specifications and to develop
the design and implementation of software as well as some hierarchy of
mathematical structures,
while proving that implementations (i.e. definitions) meet their specifications or design
requirements (i.e. the properties that they are supposed to satisfy).
Moreover, the object-oriented
features of this language enable the development of an implementation
by iterative refinement of its specification: 
many software
components implemented can be built by
inheritance and parameterization from already defined components.


\section{From binary relations to functions}

In this section, we present a small development illustrating how
\focalize can be used to teach basic notions on binary relations and
functions and how, at the same time, 
some knowledge on F-IDE usage can be introduced.
To validate our approach we simultaneously introduce concepts involved
in \focalize and discrete mathematics.

\paragraph{Specification of binary relations}
In FoCaLiZe, the primitive entity of a development is the \emph{species}.
Species are the nodes of the
hierarchy of structures that makes up a development.
A species can be seen as a set of ``things'', called methods, related
to the same concept. As in most modular design systems
(i.e. object-oriented, abstract data types, etc.) the idea is
to group a data structure with the operations on the data structure, the
specification of these operations (in the form of properties), the
representation requirements, and the proofs of the properties.
Therefore there are three kinds of methods: the carrier type, the
programming methods which are functions  and the logical methods which
are statements, called here properties,  and proofs.
Each method is identified by
its name and can be either declared 
(primitive constants, operations and properties)
or defined (implementation of operations, proofs of theorems).

In discrete mathematics, objects are often defined at an abstract
level. For example, a binary relation $R$ is generally defined as a subset
of a cartesian product $A \times B$. In fact, to define a relation we first need
two sets $A$ and $B$ from which the relation can be built: we don't know anything about these sets
but we have to be able to manipulate their elements to describe
elements belonging to~$R$. Hence, the
species {\footnotesize \tt Binary\_relations} 
of binary relations is parameterized by the sets {\footnotesize \tt $A$:Setoid} 
and~{\footnotesize \tt $B$:Setoid}  (where the species {\footnotesize \tt Setoid} specifies
non-empty sets together with an equivalence relation {\footnotesize \tt equal}, 
see~table~\ref{spec_setoid}).

Indeed,
an important feature of \focalize is the ability to
parameterize a species by generic collections instanciating a
species. Such a
mechanism allows us to use a species, 
without embedding its methods (which is the inheritance mechanism) in the new
structure but to use it as a tool box to build this new structure  
by calling its methods
explicitly  without knowing how the methods --- the tools --- are built.

Each species must have one unique carrier method, or representation
type: it corresponds to
the concrete representation of
the elements of the set underlying the structure defined
by the species. 
The carrier is represented by the keyword {\footnotesize \tt Self} inside
the species and outside, by the name of the species itself, so that we
identify the
set with the structure, as usual in mathematics.
Like all the other methods, the carrier
can be either declared or defined. A declared carrier 
denotes any set (as in the sentence ``let $E$ be a set''), 
while a defined one is a binding to a concrete type. 

In the species {\footnotesize \tt Binary\_relations}, nothing is said about
how to implement relations and the carrier method {\footnotesize \tt Self} is only
declared: we write {\footnotesize \tt $R$:Self} to express that $R$ is a
relation belonging to the species
{\footnotesize \tt Binary\_relations}. In this context, we are now in
position to specify what is a binary relation by introducing a method
{\footnotesize \tt relation: Self -> $A$ -> $B$ -> bool}
corresponding to characteristic
functions of relations (given a relation {\footnotesize \tt $R$:Self}, for
$a:A$ and $b:B$, 
{\footnotesize \tt relation($R,a,b$)=true} iff  $(a,b) \in R$). At this level of the hierarchy,
the method {\footnotesize \tt relation} is only declared (we don't describe 
particular relations but only what is needed to define a relation).

Another important feature of \focalize is the
inheritance
mechanism: one can
enrich a species with additional operations (methods) and
redefine some methods of the parent species, but one can also get
closer to a runnable implementation by providing explicit definitions to methods
that were only declared in the parents.
A species can
inherit the declarations and definitions of one or several already
defined species and is
free to define or
redefine an inherited method as long as such (re)definition does not
change the type of the method.

For example, in mathematics,  the set of binary relations is endowed with a notion of
equality derived from the equalities of the two component sets.  This
equality turns this set of binary relations into a setoid. We can
easily express that point  by indicating that  the species {\footnotesize \tt Binary\_relations}
inherits from the species {\footnotesize \tt Setoid}. 
In this way, in
{\footnotesize \tt Binary\_relations} and in all species inheriting
from it, the method 
{\footnotesize \tt equal} can be called to compare
relations. Moreover, since the parameters $A$ and $B$ are also
setoids, the syntactic construction {\footnotesize \tt A!equal}
(resp. {\footnotesize \tt B!equal}) can be used to call the method
{\footnotesize \tt equal} of the species $A$ (resp. $B$) to compare elements of $A$ (resp. of~$B$).

\begin{center}
\begin{scriptsize}
\begin{tabular}{|l|}
\hline
{\tt species Binary\_relations (A is Setoid, B is Setoid) =} \\
{\tt inherit Setoid;} \\
{\tt signature relation : Self -> A -> B -> bool;}\\
{\tt end ;;}\\
\hline
\end{tabular}
\end{scriptsize}
\end{center}

\noindent
Of course, (we hope that) many students know what is a binary relation. However,
here, introducing the species of binary relations leads to introduce
(at a very basic level)
computer science concepts such as parameters, inheritance, abstract
and concrete data types, declarations and definitions.

\paragraph{Specifications, Definitions and Proofs}
At this point, the
method {\footnotesize \tt equal} is only declared in the species
{\footnotesize \tt Setoid}
and it remains to define it 
in the species {\footnotesize \tt Binary\_relations}. To achieve this
goal, we can declare the method {\footnotesize \tt is\_contained $\colon$ Self -> Self -> bool} 
such that
{\footnotesize \tt is\_contained($R_1,R_2$)=true} iff $R_1 \subseteq
R_2$. Hence, we add the signature of {\footnotesize \tt is\_contained} together with a property 
expressing the specification
of this method in the species {\footnotesize \tt Binary\_relations}.

\begin{center}
\begin{scriptsize}
\begin{tabular}{|l|}
\hline
{\tt signature is\_contained: Self -> Self -> bool ;}\\
{\tt property is\_contained\_spec: all r1 r2: Self,}\\ 
{\tt  \hspace{0.2cm} is\_contained(r1, r2) <->  all a: A, all b: B, relation(r1, a, b) -> relation(r2, a, b);}\\
\hline
\end{tabular}
\end{scriptsize}
\end{center}

\noindent
Declared methods are introduced by the keyword
{\footnotesize \tt signature} while defined methods are introduced by {\footnotesize \tt let} and recursive
definitions must be explicitely flagged with the keyword {\footnotesize \tt rec}.
The method {\footnotesize \tt is\_contained\_spec} corresponds to a
logical method. 
Such methods represent the properties of
programming methods. The declaration of a logical method 
is simply the statement of a property, while the definition is a proof of
this statement. In the first case, we speak of {properties} ({\footnotesize
  \tt property}) that are still
to be proved later in the development, while in the second case we speak of
{theorems} ({\footnotesize \tt theorem}). 
The language also permits logical definitions ({\footnotesize \tt logical
  let}) to bind names to logical statements.
The language used for the statements is composed of the basic logical
connectors {\footnotesize \verb+\/+}, {\footnotesize \verb+/\+}, 
{\footnotesize \tt ->}, {\footnotesize \tt <->}, {\footnotesize \tt not}, 
and universal ({\footnotesize \tt all})
and existential ({\footnotesize \tt ex}) quantification over a \focalize type.

As we can see in our example,
as usual during a formal development (and as required as a good
practice when applying formal methods), specifications are provided
before implementations. Later, during inheritance,
the method 
{\footnotesize \tt is\_contained} will have to be implemented and the proof of 
{\footnotesize \tt is\_contained\_spec} will have to be done.
However, even if this method is only declared, it is possible to use it in a
definition. For example, we can now define the method 
{\footnotesize \tt equal} (which is still only declared) over
relations and we can prove the required properties on this definition 
(as specified in the species {\footnotesize \tt Setoid}, this method must define a reflexive, symmetric and transitive
relation over {\footnotesize \tt Self}).

\begin{center}
\begin{scriptsize}
\begin{tabular}{|l|}
\hline
{\tt let equal(x, y) = is\_contained(x, y) \&\& is\_contained(y, x) ;}\\
{\tt theorem equal\_spec : all r1 r2 : Self,} \\
{\tt \hspace{0.2cm} equal (r1, r2) <-> (all a : A, all b : B, relation(r1, a, b) <-> relation (r2, a, b))} \\
{\tt \hspace{0.2cm}  proof = by definition of equal} \\
{\tt \hspace{2.1cm} property is\_contained\_spec ;}\\
{\tt proof of equal\_reflexive = by property equal\_spec;}\\
{\tt proof of equal\_symmetric = by property equal\_spec;}\\
{\tt proof of equal\_transitive = by property equal\_spec;}\\
\hline
\end{tabular}
\end{scriptsize}
\end{center}

\noindent
In fact, the method  {\footnotesize \tt equal} is defined
together with a proved theorem  {\footnotesize \tt equal\_spec} corresponding to its specification.
The proof is obtained in an automatic way: we just specify here that
it can be done by considering the definition of {\footnotesize \tt equal} 
and the specification {\footnotesize \tt is\_contained\_spec} (we
don't specify how these methods have to be used to make the proof).
Thanks
to this theorem, proofs of reflexivity, symmetry and transitivity of 
{\footnotesize \tt equal} are obvious and can also be automatically done
(it suffices to indicate that they can be obtained by considering the
theorem {\footnotesize \tt equal\_spec}). 

There are no difficulties to do such mathematical proofs, which can
be more detailed if needed to point out the mathematical reasonment. 
Now, there is, on the computer science side,  a question which naturally arises from this tiny
development. What is the consequence of redefining the equality in a
species inheriting from {\footnotesize \tt Binary\_relations}? Any
proof relying on the definition of {\footnotesize \tt equal} should be
redone (and the compiler leaves no room to an attempt to keep the old
version). This is the time to try another version by directly using
the definition of {\footnotesize \tt equal} to prove reflexivity,
symmetry and transitivity and to find out that these proofs have to be
invalidated when redefining  {\footnotesize \tt equal}. This puts the
emphasis on the benefit obtained from the introduction of the
specification of {\footnotesize \tt equal}: only the proof of 
{\footnotesize \tt equal\_spec} is to be redone in case of
redefinition of {\footnotesize \tt equal} while the proofs 
of reflexivity, symmetry and transitivity remain valid since they do
not depend on the definition of {\footnotesize \tt equal}. Hence, it
is demonstrated that,  to minimize the impact of
redefinitions,  proofs must  rely on specification properties instead on
definitions (this point is discussed in~\cite{calc03}).

Therefore,
as we can see here,
even in a very simple and small example on discrete mathematics, some non-trivial methodological
issues in computer science can be addressed.

\begin{table}
\begin{center}
\begin{scriptsize}
\begin{tabular}{|l|}
\hline
{\verb+species Setoid =+} \\
{\verb+ inherit Basic_object;+} \\
{\verb+ signature element : Self;+}\\
{\verb+ signature equal : Self -> Self -> bool;+} \\
{\verb+ property equal_reflexive : all x : Self, equal(x,x);+}\\
{\verb+ property equal_symmetric : all x y : Self, equal(x,y) -> equal(y,x);+}\\
{\verb+ property equal_transitive : all x y z : Self, equal(x,y) -> equal(y,z) -> equal(x,z);}+}\\
{\verb+ let different (x, y) = not (equal(x,y));+}\\
{\verb+ theorem same_is_not_different : all x y : Self, different(x,y) <-> not (equal(x,y))+}\\
{\verb+  proof = by definition of different;+}\\
{\verb+end;;+}\\
\hline
\end{tabular}
\end{scriptsize}
\end{center}
\caption{Species of setoids}\label{spec_setoid}
\end{table}

\paragraph{Formal reasoning on mathematical properties}
At an abstract level, \focalize allows us to introduce 
some properties. For example, in the context of a discrete mathematics
course, one can define what is an injective relation, a surjective
relation, a deterministic relation and a left-total relation by adding
the following methods in the species  {\footnotesize \tt Binary\_relations}.

\begin{center}
\begin{scriptsize}
\begin{tabular}{|l|}
\hline
\verb+logical final let is_left_unique(r) = all a1 a2 : A, all b : B,+\\
\verb+            (relation(r,a1,b) /\ relation(r,a2,b)) -> A!equal(a1,a2);+\\
\verb+logical final let is_right_total(r) = all b : B, ex a : A, relation(r,a,b);+\\
\verb+logical final let is_right_unique(r) = all a : A, all b1 b2 : B,+\\
\verb+            (relation(r,a,b1) /\ relation(r,a,b2)) -> B!equal(b1,b2);+\\
\verb+logical final let is_left_total(r) = all a : A, ex b : B, relation(r, a, b);+ \\
\hline
\end{tabular}
\end{scriptsize}
\end{center}

\noindent
These methods correspond to  definitions of logical properties: they only
bind names 
to statements and
don't intend to express that these properties are true or false (contrarily to the
methods introduced by {\footnotesize \tt property}).
The keyword {\footnotesize \tt final} is used to forbid the
redefinition of these methods in the species inheriting from {\footnotesize \tt Binary\_relations}.
We can also describe the empty relation, the full relation, and
singleton relations as follows.

\begin{center}
\begin{scriptsize}
\begin{tabular}{|l|}
\hline
\verb+logical final let is_empty_r(r) = all a : A, all b : B, not relation(r,a,b) ;+\\
\verb+logical final let is_full_r(r) = all a : A, all b : B, relation(r,a,b) ;+\\
\verb+logical final let is_singleton_r(r,a,b) = all a1 : A, all b1 : B, +\\
\verb+            relation(r,a1,b1) <-> (A!equal(a,a1) /\ B!equal(b,b1));+\\
\hline
\end{tabular}
\end{scriptsize}
\end{center}

\noindent
Similarly, we can introduce
operations by only
specifying their properties (like in logic programming languages). For
example, we can describe union, intersection and difference of
relations as follows.

\begin{center}
\begin{scriptsize}
\begin{tabular}{|l|}
\hline
\verb+logical final let is_union_r(r1,r2,r3) = all a : A, all b : B,+\\ 
\verb+            relation(r3,a,b) <-> (relation(r1,a,b) \/ relation(r2,a,b));+\\
\verb+logical final let is_intersection_r(r1,r2,r3) = all a : A, all b : B, +\\
\verb+            relation(r3,a,b) <-> (relation(r1,a,b) /\ relation(r2,a,b));+\\
\verb+logical final let is_diff_r(r1,r2,r3) = all a : A, all b : B, +\\
\verb+            relation(r3,a,b) <-> (relation(r1,a,b) /\ not relation(r2,a,b));+\\
\hline
\end{tabular}
\end{scriptsize}
\end{center}

\noindent
Thanks to these methods, it becomes possible to prove classical
properties, often done as exercices during discrete mathematics
courses. For example we can prove the following property.
\[
\left (
\begin{array}{l}
R_1 \, \, \mbox{is injective}
\land 
R_2 \, \, \mbox{is injective} \\
\land \, \forall a_1,a_2:A \, \forall b:B \, \, ((a_1,b) \in R_1 \land
(a_2,b) \in R_2) \Rightarrow a_1=a_2
\end{array}
\right ) \Leftrightarrow
R_1 \cup R_2 \, \, \mbox{is injective} 
\]
In the context of a discrete mathematics course, the goal is
not here to make the proofs with the automatic features of Zenon but to
write a detailed proof of a mathematical property. Hence, we would
like to formally prove the following theorem.

\begin{center}
\begin{scriptsize}
\begin{tabular}{|l|}
\hline
{\tt theorem union\_is\_left\_unique : all r1 r2 r3 : Self,}\\
{\tt \hspace{0.1cm} is\_union\_r(r1,r2,r3) ->}\\
{\tt \hspace{0.2cm}  ((is\_left\_unique(r1) \verb+/\+ is\_left\_unique(r2)}\\
{\tt  \hspace{0.2cm}  \verb+/\ (all a1 a2: A, all b: B,((relation (r1,a1,b) /\ relation(r2,a2,b)) -> A!equal(a1,a2))))+}\\
{\tt  \hspace{0.2cm}   <-> is\_left\_unique(r3))}\\
\hline
\end{tabular}
\end{scriptsize}
\end{center}

\noindent
Within FoCaLiZe,
a proof is a tree where the programmer
introduces names ({\footnotesize \tt assume}) and hypotheses ({\footnotesize \tt hypothesis}), gives a statement to
prove ({\footnotesize \tt prove}) and then provides justification for the
statement. This justification can be: (1) a ``{\footnotesize \tt conclude}'' clause for
fully automatic proof; (2) a ``{\footnotesize \tt by}'' clause with a list of
definitions, properties, hypotheses, previous theorems, and previous
steps (subject to some scoping conditions) for use by the automatic
prover; (3) a sequence of proofs (with their own assumptions,
statements, and proofs) whose statements will be used by the automatic
prover to prove the current statement.
Hence, each step of a proof is independent of the others and can be
reused in a similar context\footnote{This eases maintenance of
  proofs, and allows us to
use exactly the same proof for a statement based on an
hypothesis $A$ and for the same statement based on a stronger
hypothesis $B$, provided the automatic prover can make the inference
from $B$ to~$A$.}. Thanks to these features, as illustrated in
table~\ref{proof_inj}, a formal proof (left side of table~\ref{proof_inj}), very close to the
informal proof (right side of table~\ref{proof_inj}), of the theorem 
{\footnotesize \tt union\_is\_left\_unique} can be done within
FoCaLiZe.
As we can see, the structure of the proof appears clearly (proving an
equivalence leads to prove two implications, proving an implication
leads to assume hypothesis and to prove the conclusion, proving a
conjunction leads to prove each member of the conjunction, using an
implication to prove a statement leads to prove hypothesis of this
implication, etc.) and each step is clearly characterized by some
assumptions and a goal to prove. 
Hence, using \focalize during a mathematics course can guide students when
specifying  and proving classical properties by providing some help to
answer questions: is this specification correct according to this
property~? are these properties needed to prove this statement~? is there 
an implicit assumption in this proof ? is this statement provable by
using these proof steps~? etc.

\begin{table}
\begin{center}
{\fontsize{7}{0.7em} \selectfont
\begin{tabular}{|ll|}
\hline
{\tt proof =} &\\
\hspace{0.2cm}  \verb+<0>1 assume r1 r2 r3 : Self,+ & Let
$R_1$, $R_2$ and $R_3$ be binary relations.\\
\hspace{1cm}  \verb+hypothesis Hunion : is_union_r(r1,r2,r3),+ &
such that $R_3 = R_1 \cup R_2$.\\
\hspace{1cm} \verb+prove (is_left_unique(r1) /\ is_left_unique(r2)+ &
Let us prove the desired equivalence.\\
\hspace{1cm}  \verb+       /\(all a1 a2 : A, all b : B,+ & \\
\hspace{1cm}  \verb+          ((relation(r1,a1,b) /\ relation(r2,a2,b))+ & \\
\hspace{1cm}  \verb+           -> A!equal (a1, a2))))+ & \\
\hspace{1cm}  \verb+     <-> is_left_unique(r3)+ & \\
\mbox{} & \\
\hspace{0.5cm}  \verb+<1>1 hypothesis Hlu1 : is_left_unique(r1),+ &
First, let us suppose that $R_1$ is injective, \\
\hspace{1.4cm}  \verb+hypothesis Hlu2 : is_left_unique(r2),+ & $R_2$
is injective, \\
\hspace{1.4cm}  \verb+hypothesis Heq : all a1 a2 : A, all b : B,+& 
and that $\forall a_1,a_2:A \, \forall b:B$ \\
\hspace{1.4cm}  \verb+      ((relation (r1,a1,b) /\ relation(r2,a2,b))+& 
$((a_1,b) \in R_1 \land (a_2,b) \in R_2) \Rightarrow a_1=a_2$\\
\hspace{1.4cm}  \verb+      -> A!equal(a1,a2)),+ & \\
\hspace{1.4cm}  \verb+prove is_left_unique(r3)+ & and let us prove
that $R_3$ is injective.\\
\hspace{0.8cm}  \verb+<2>1 assume a1 a2 : A, assume b : B,+ & Let
$a_1,a_2:A$ and $b:B$ be elements\\
\hspace{1.6cm}  \verb+hypothesis Ha1 : relation(r3,a1,b),+ &
such that $(a_1,b) \in R_3$ \\
\hspace{1.6cm}  \verb+hypothesis Ha2 : relation(r3,a2,b),+ &
and $(a_2,b) \in R_3$, \\
\hspace{1.6cm}  \verb+prove A!equal(a1, a2)+& and let us prove that
$a_1=a_2$.\\
& We consider 4 cases.\\
\hspace{1.1cm}  \verb+<3>1 hypothesis H11 : relation(r1,a1,b),+ & If
we suppose that $(a_1,b) \in R_1$,\\
\hspace{1.9cm}  \verb+hypothesis H12 : relation(r1,a2,b),+ & and $(a_2,b) \in R_1$,\\
\hspace{1.9cm}  \verb+prove A!equal(a1,a2)+ & then we can prove
$a_1=a_2$ \\
\hspace{1.9cm}  \verb+by hypothesis H11, H12, Hlu1+ & since (by
hypothesis) $R_1$ is injective,\\
\hspace{1.9cm}  \verb+   definition of is_left_unique+ & and by
definition of an injective relation.\\
\hspace{1.1cm}  \verb+<3>2 hypothesis H21 : relation(r2,a1,b),+ & If
we suppose that $(a_1,b) \in R_2$,\\
\hspace{1.9cm}  \verb+hypothesis H22 : relation(r2,a2,b),+ & and $(a_2,b) \in R_2$,\\
\hspace{1.9cm}  \verb+prove A!equal(a1,a2)+ & then we can prove
$a_1=a_2$ \\
\hspace{1.9cm}  \verb+by hypothesis H21, H22, Hlu2+ & since (by
hypothesis) $R_2$ is injective,\\
\hspace{1.9cm}  \verb+   definition of is_left_unique+ & and by
definition of an injective relation.\\
\hspace{1.1cm}  \verb+<3>3 hypothesis H31 : relation(r1,a1,b),+ & If
we suppose that $(a_1,b) \in R_1$,\\
\hspace{1.9cm}  \verb+hypothesis H32 : relation(r2,a2,b),+ & and $(a_2,b) \in R_2$,\\
\hspace{1.9cm}  \verb+prove A!equal(a1,a2)+ & then we can prove
$a_1=a_2$ \\
\hspace{1.9cm}  \verb+by hypothesis H31, H32, Heq+ & by using
hypothesis (\verb+Heq+). \\ 
\hspace{1.1cm}  \verb+<3>4 hypothesis H41 : relation(r2,a1,b),+ & If
we suppose that $(a_1,b) \in R_2$,\\
\hspace{1.9cm}  \verb+hypothesis H42 : relation(r1,a2,b),+ & and $(a_2,b) \in R_1$,\\
\hspace{1.9cm}  \verb+prove A!equal(a1,a2)+ & then we can prove
$a_1=a_2$ \\
\hspace{1.9cm}  \verb+by hypothesis H41, H42, Heq+ & by using
hypothesis (\verb+Heq+). \\ 
\hspace{1.1cm}  \verb+<3>f qed by step <3>1, <3>2, <3>3, <3>4+ & In
these 4 cases, $a_1=a_2$ and \\
\hspace{1.1cm}  \verb+            hypothesis Hunion,+ & since
by hypothesis $R_3 = R_1 \cup R_2$, \\
\hspace{1.1cm}  \verb+                       Ha1, Ha2+ & and
 $(a_1,b) \in R_3$ and $(a_2,b) \in R_3$, \\
\hspace{1.1cm}  \verb+            definition of is_union_r+ & we can
conclude by definition of $\cup$.\\
\mbox{} & \\
\hspace{0.5cm}  \verb+<1>2 hypothesis Hlu3 : is_left_unique(r3),+ &
Now, let us suppose that $R_3$ is injective, \\
\hspace{1.2cm}  \verb+prove is_left_unique(r1) /\ is_left_unique(r2)+
& and let us prove that $R_1$ and $R_2$ are injective,  \\
\hspace{1.2cm}  \verb+/\ (all a1 a2 : A, all b : B,+& and are such that
$\forall a_1,a_2:A$  $\forall b:B$, \\
\hspace{1.2cm}\verb+((relation(r1,a1,b)/\relation(r2,a2,b))->A!equal(a1,a2)))+ &
$((a_1,b) \in R_1 \land (a_2,b) \in R_2) \Rightarrow a_1=a_2$ 
\\
\hspace{0.8cm}  \verb+<2>1 prove is_left_unique(r1)+ & We first prove
that $R_1$ is injective.\\
\hspace{1.1cm}  \verb+<3>1 assume a1 a2 : A, assume b : B,+ & Let
$a_1,a_2:A$ and $b:B$ be elements\\
\hspace{1.9cm}  \verb+hypothesis Hr1:relation(r1,a1,b)/\relation(r1,a2,b),+ &
such that $(a_1,b) \in R_1 \land (a_2,b) \in R_1$,\\
\hspace{1.9cm}  \verb+prove A!equal(a1,a2)+ &
and let us prove that $a_1=a_2$.\\
\hspace{1.4cm}  \verb+<4>1 prove relation(r3,a1,b) /\ relation(r3,a2,b)+ &
We prove that $(a_1,b) \in R_3 \land (a_2,b) \in R_3$ \\ 
\hspace{2.2cm}  \verb+by hypothesis Hr1, Hunion definition of is_union_r+ &
since $R_3 = R_1 \cup R_2$ and by definition of $\cup$.\\
\hspace{1.4cm}  \verb+<4>f qed by step <4>1 hypothesis Hlu3+ & Hence,
since $R_3$ is injective, we get $a_1=a_2$\\
\hspace{1.4cm}  \verb+            definition of is_left_unique+
& by definition of an injective relation.\\
\hspace{1.1cm}  \verb+<3>f qed by step <3>1 definition of is_left_unique+ & 
Thus, by definition, $R_1$ is also injective. \\
\hspace{0.8cm}  \verb+<2>2 prove is_left_unique(r2)+ & Similarly we prove
that $R_2$ is injective.\\
\hspace{1.1cm}  \verb+<3>1 assume a1 a2 : A, assume b : B,+ & Let
$a_1,a_2:A$ and $b:B$ be elements\\
\hspace{1.9cm}  \verb+hypothesis Hr2:relation(r2,a1,b)/\relation(r2,a2,b),+ &
such that $(a_1,b) \in R_2 \land (a_2,b) \in R_2$,\\
\hspace{1.9cm}  \verb+prove A!equal(a1,a2)+ &
and let us prove that $a_1=a_2$.\\
\hspace{1.4cm}  \verb+<4>1 prove relation(r3,a1,b) /\ relation(r3,a2,b)+ &
We prove that $(a_1,b) \in R_3 \land (a_2,b) \in R_3$ \\ 
\hspace{2.2cm}  \verb+by hypothesis Hr2, Hunion definition of is_union_r+ &
since $R_3 = R_1 \cup R_2$ and by definition of $\cup$.\\
\hspace{1.4cm}  \verb+<4>f qed by step <4>1 hypothesis Hlu3+ &
Hence,since $R_3$ is injective, we get $a_1=a_2$ \\
\hspace{1.4cm}  \verb+            definition of is_left_unique+
& by definition of an injective relation.\\
\hspace{1.1cm}  \verb+<3>f qed by step <3>1 definition of is_left_unique+ & 
Thus, by definition, $R_2$ is also injective. \\
\hspace{0.8cm}  \verb+<2>3 prove all a1 a2 : A, all b : B,+& It
remains to prove that \\
\hspace{0.8cm}  \verb+         ((relation(r1,a1,b) /\ relation(r2,a2,b))+ 
&  $\forall a_1,a_2:A \, \forall b:B \, \, ((a_1,b) \in R_1 \land
(a_2,b) \in R_2) $ \\
\hspace{0.8cm}  \verb+         -> A!equal(a1,a2))+ & \hspace{1.8cm} $\Rightarrow a_1=a_2$\\
\hspace{1.1cm}  \verb+<3>1 assume a1 a2 : A, assume b : B,+ & Let
$a_1,a_2:A$ and $b:B$ be elements \\
\hspace{1.1cm}  \verb+     hypothesis H0:relation(r1,a1,b)/\relation(r2,a2,b),+
& such that $(a_1,b) \in R_1 \land (a_2,b) \in R_2$.\\
\hspace{1.1cm}  \verb+     prove relation(r3,a1,b) /\ relation(r3,a2,b)+ &
We can prove that $(a_1,b) \in R_3 \land (a_2,b) \in R_3$\\
\hspace{1.1cm}  \verb+     by hypothesis H0, Hunion definition of is_union_r+ &
since  $R_3 = R_1 \cup R_2$ and by definition of $\cup$.\\
\hspace{1.1cm}  \verb+<3>f qed by step <3>1 hypothesis Hlu3+
& Hence,since $R_3$ is injective, we get $a_1=a_2$  \\
\hspace{1.1cm}  \verb+            definition of is_left_unique+
& by definition of an injective relation. \\
\hspace{0.8cm}  \verb+<2>f conclude+ & This concludes the proof of the
conjunction \verb+<2>1+. \\
\hspace{0.5cm}  \verb+<1>f conclude+ & This concludes the proof of the
equivalence \verb+<0>1+.\\
\mbox{} & \\
\hspace{0.2cm}  \verb+<0>f conclude;+ & This concludes the proof of
the theorem. \\
\hline
\end{tabular}
}
\end{center}
\caption{Proof of theorem {\footnotesize \tt union\_is\_left\_unique}}\label{proof_inj}
\end{table}

\paragraph{Building a hierarchy of mathematical structures}
Adding the specifications of operations over relations and the
classical properties over relations in the species {\footnotesize \tt Binary\_relations} only 
leads to bind names to properties without
asserting if these properties are true or false.
It is now possible to build a hierarchy of species inheriting from
{\footnotesize \tt Binary\_relations} in order to constrain relations
to satisfy some of these properties. For example, the species of
injective relations can be introduced as follows (we just consider
here one theorem to illustrate exercices that can be done at this level).

\begin{center}
\begin{scriptsize}
\begin{tabular}{|l|}
\hline
\verb+species Injective_relations(A is Setoid, B is Setoid) =+\\
\verb+ inherit Binary_relations(A, B);+\\
\verb+ property left_unique : all r : Self, is_left_unique(r);+\\
\verb+ theorem injective_union : all r1 r2 r3 : Self, +\\
\verb+    is_union_r(r1,r2,r3) +\\
\verb+    -> (all a1 a2: A, all b: B,((relation(r1,a1,b) /\ relation(r2,a2,b))->A!equal(a1,a2)))+\\
\verb+  proof = by property left_unique, union_is_left_unique ;+\\
\verb+end;;+\\
\hline
\end{tabular}
\end{scriptsize}
\end{center}

\noindent
Here, we can use the theorem {\footnotesize \tt union\_is\_left\_unique} 
and the property {\footnotesize \tt left\_unique} (necessarily satisfied by all
elements of type {\footnotesize \tt Self}) to prove properties over
union of relations (note that the definition of {\footnotesize \tt is\_left\_unique} 
is not used in this proof which is obtained by only considering
properties of logical connectors between statements). 
This can also be done for all the operations and properties previously
introduced. 
For example, we can introduce the species of deterministic and
left-total relations as follows.

\begin{center}
\begin{scriptsize}
\begin{tabular}{|l|}
\hline
\verb+species Deterministic_relations (A is Setoid, B is Setoid) =+\\
\verb+ inherit Binary_relations (A, B);+\\
\verb+  property right_unique : all r : Self, is_right_unique (r);+\\
\verb+end;;+\\
\verb+species Left_total_relations (A is Setoid, B is Setoid) =+\\
\verb+ inherit Binary_relations (A, B);+\\
\verb+ property left_total : all r : Self, is_left_total (r);+\\
\verb+end;;+\\
\hline
\end{tabular}
\end{scriptsize}
\end{center}

\noindent
Furthermore, we can go one step further and build a
``complete'' hierarchy by considering functions, injective functions,
surjective functions and bijective functions as particular cases of
relations.
This leads to build the following hierarchy of relations corresponding
to usual contents in a mathematics course.

\begin{center}
\begin{footnotesize}
\begin{tikzpicture}[thick,>=stealth',auto]
\tikzstyle{gris}=[fill=gray!40]

\node (BR) at (0,0)  {\fbox{Binary Relations}};
\node (PBR) at (0,-0.15)  {};

\node (IR) at (-4.5,-1.25)  {\fbox{\begin{tabular}{c} Injective \\
      Relations \end{tabular}}};

\node (DR) at (-1.5,-1.25)  {\fbox{\begin{tabular}{c} Deterministic \\
      Relations \end{tabular}}};

\node (LR) at (1.5,-1.25)  {\fbox{\begin{tabular}{c} Left total \\
      Relations \end{tabular}}};

\node (SR) at (4.5,-1.25)  {\fbox{\begin{tabular}{c} Surjective \\
      Relations \end{tabular}}};

\node (F) at (0,-2.5)  {\fbox{Functions}};

\node (IF) at (-3,-2.75)  {\fbox{Injective Functions}};

\node (SF) at (3,-2.75)  {\fbox{Surjective Functions}};

\node (BF) at (0,-3.75)  {\fbox{Bijective Functions}};

\path[->,bend right] (BR) edge (IR); 
\draw[->,bend left] (BR) to (SR); 
\path[->] (BR) edge (LR); 
\path[->] (BR) edge (DR); 
\path[->] (DR) edge (F); 
\path[->] (LR) edge (F); 
\path[->] (IR) edge (IF); 
\path[->] (F) edge (IF); 
\path[->] (SR) edge (SF); 
\path[->] (F) edge (SF); 
\path[->] (IF) edge (BF); 
\path[->] (SF) edge (BF);

\end{tikzpicture}
\end{footnotesize}
\end{center}

\noindent
However, in the species  {\footnotesize \tt Functional\_relations} of
functions (and
in all the species inheriting from it), elements of type {\footnotesize \tt Self} are still defined by
their characterisitic functions {\footnotesize \tt relation}: this
leads to view functions from $A$ to $B$ as 
particular cases of relations over $A \times B$. 
However, it may be useful to declare a method 
{\footnotesize \verb+fct : Self -> A -> B+} corresponding to the usual
concept of functions (known by students at the undergraduate level).
From this method, it becomes possible to define the method 
{\footnotesize \tt relation} and to prove the required properties. 
This can be easily done as follows.

\begin{center}
\begin{scriptsize}
\begin{tabular}{|l|}
\hline
\verb+species Functional_relations (A is Setoid, B is Setoid) =+\\
\verb+ inherit Left_total_relations(A, B), Deterministic_relations(A, B);+\\
\verb+ signature fct : Self -> A -> B;+\\
\verb+ let relation(r,x,y) = B!equal(fct(r,x),y);+\\
\verb+ proof of right_unique = by definition of relation, is_right_unique+\\
\verb+                            property B!equal_symmetric, B!equal_transitive;+\\
\verb+ proof of left_total = by definition of relation, is_left_total property B!equal_reflexive;+\\
\verb+end;;+\\
\hline
\end{tabular}
\end{scriptsize}
\end{center}

\noindent
In addition to the mathematical contents of these specifications
(allowing students to understand at a deep
level the differences between functions and relations, and the
main properties these objects),
using
\focalize to describe the hierarchy of relations and functions allows
students to consider multiple-inheritance and computational notions.

\paragraph{Implementations and their properties}
Until now, we have only used \focalize to express, to prove and to
design the architecture of mathematical properties. The next step
consists in introducing
concrete data types,
recursive programming and inductive proofs over mathematical objects.
We show here how  to introduce
these notions by implementing finite parts of a set by lists. We
first define the species 
(parameterized by a setoid {\footnotesize \tt S})
of finite parts of 
{\footnotesize \tt S} (due to space limitations, we only consider the methods needed
in our example, but, of course, this species contains many other methods).

\begin{center}
\begin{scriptsize}
\begin{tabular}{|l|}
\hline
\verb+species Finite_parts(S is Setoid) =+\\
\verb+ inherit Setoid ;+\\
\verb+ signature belongs: S -> Self -> bool;+\\
\verb+ signature cardinal: Self -> int;+\\
\verb+ signature empty : Self;+\\
\verb+ signature release : Self -> S -> Self;+\\
\verb+ property release_spec : all x : Self, all t1 t2 : S,+\\
\verb+    belongs(t1,release(x,t2)) <-> (S!different(t1,t2) /\ belongs(t1,x));+\\
\verb+ property empty_spec :all t: S, not belongs(t,empty);+\\
\verb+ signature from_list : list(S) -> Self ;+\\
\verb+ property belongs_spec: all t: list(S), all h x: S,+\\
\verb+  (belongs(x,from_list(t)) \/ S!equal(h,x)) <-> belongs(x,from_list(h::t));+\\
\verb+end;;+\\
\hline
\end{tabular}
\end{scriptsize}
\end{center}

\noindent
Hence, a finite part {\footnotesize \tt $P$:Self} of  
{\footnotesize \tt S} is described by a membership relation
({\footnotesize \tt belongs($s,P$)=true} iff $s \in P$) and by its
cardinal (which is finite since it is represented here by an integer).
In our example, we consider the methods 
{\footnotesize \tt empty} (for the empty part) and
{\footnotesize \tt release}
(that permits to remove an element from a finite part). At this abstract
level, these methods are only declared together with their
specifications.
Furthermore, we declare a method {\footnotesize \tt from\_list}  that aims at building a finite part
from elements belonging to a list and which is used to specify the
method {\footnotesize \tt belong}.
We can now refine this species by representing finite parts with
the concrete \focalize type {\footnotesize \tt list} 
of lists.  
Within FoCaLiZe, the
language used for the programming methods is similar to the functional core of 
OCaml (let-binding, pattern matching, conditional,
higher order functions, etc),
with the addition of a construction to call a method from a given
structure. 
Thanks to these constructions,  
we can introduce the species {\footnotesize \verb+Finite_parts_by_lists+}
inheriting from  {\footnotesize \verb+Finite_parts+}
and providing definitions for the
programming methods.

\begin{center}
\begin{scriptsize}
\begin{tabular}{|l|}
\hline
\verb+species Finite_parts_by_lists(S is Setoid) =+\\
\verb+ inherit Finite_parts(S);+\\
\verb+ representation = list(S);+\\
\verb+ let rec belongs(x:S,l) = match l with | [] -> false+\\
\verb+    | h :: q ->  S!equal(h,x) || (belongs(x,q))+\\
\verb+ termination proof = structural l+;\\
\verb+ let rec cardinal(e) = match e with | [] -> 0+\\
\verb~    | _ ::t -> 1 + cardinal(t)~\\
\verb+ termination proof = structural e;+ \\
\verb+ let empty = [];+\\
\verb+ let rec release(e,s) = match e with | [] -> []+\\
\verb+  | h::t -> if S!equal(s,h) then release(t,s) else h::release(t,s)+\\
\verb+  termination proof = structural e ;+\\
\verb+ let from_list (s:list(S)):Self = s;+ \\
\verb+end;;+\\
\hline
\end{tabular}
\end{scriptsize}
\end{center}

\noindent
In the context of a discrete mathematics course, this leads to
introduce recursive definitions and to (lightly) address termination issues of such definitions.

We are now in position to prove all the properties 
stated in the species {\footnotesize \tt Finite\_parts}. We just
present here the proof of {\footnotesize \tt release\_spec}. The proof is
done by induction on lists, and, here again,
as we can see in table~\ref{proof_release_spec}, the formal proof is
very close to the informal one
(in the informal proof we write $e \ominus s$ instead of 
{\footnotesize \tt release(e,s)}):  the empty list case and the inductive
step are independently proved, and the properties and definitions
leading to intermediate results are made explicit, as well as the
context in which such results are proved. Each method of the species  
{\footnotesize \tt Finite\_parts\_by\_lists}
is now defined.

\begin{table}
\begin{center}
{\fontsize{6.5}{0.6em} \selectfont
\begin{tabular}{|ll|}
\hline
\verb+proof of release_spec = + & \\
\verb+<0>1 assume e1 e2 : S,+ & Let $e_1,e_2$ be elements of $S$, and \\
\verb+     prove all l:list(S),belongs(e1,release(l,e2))+ & and let us
prove (by induction on $l$) \\
\verb+     <-> (S!different (e1,e2) /\belongs(e1,l))+ &
the desired equivalence. \\
\verb+ <1>b prove belongs(e1,release([],e2))+ & First, let us prove the
property \\
\verb+      <-> (S!different(e1, e2) /\ belongs(e1,[]))+& for the
empty list.\\
\verb+  <2>1 prove not (belongs(e1, release([], e2)))+& Since $e_1 \not
\in [ \, ] \ominus e_2 = [ \, ]$ \\
\verb+       by definition of release, empty+&
by definition of $\ominus$ and  $[ \, ]$\\
\verb+          property empty_spec+&
(because $\forall e:S \, \, e \not \in [ \, ]$)\\
\verb+  <2>2 prove not (S!different (e1,e2) /\ belongs(e1,[]))+&
and since $\neg (e_1 \neq e_2 \land e_1 \in [ \, ])$\\
\verb+       by definition of empty property empty_spec+ & 
(by definition of $[\,]$ and because $\forall e:S \, \, e \not \in [ \, ]$)\\
\verb+  <2>f conclude+& we can conclude.\\
\verb+ <1>i assume t: list(S), assume h: S,+ & For the inductive step,
let $t$ be a list and $h:S$. \\
\verb+      hypothesis HI : (belongs(e1, release(t, e2))+ &
By induction hypothesis, we have:\\
\verb+        <-> (S!different(e1,e2) /\ belongs(e1,t))),+&
$e_1 \in t \ominus e_2 \Leftrightarrow (e_1 \neq e_2 \land e_1 \in t)$,\\
\verb+      prove (belongs(e1,release(h::t,e2))+&
and we prove that\\
\verb+        <-> (S!different(e1,e2) /\ belongs(e1,h::t)))+ &
$e_1 \in h::t \ominus e_2 \Leftrightarrow (e_1 \neq e_2 \land e_1 \in h::t)$\\
\verb+  <2>1 hypothesis H1 : belongs(e1,release(h::t,e2)),+ &
($\Rightarrow$) Let us suppose that $e_1 \in h::t \ominus e_2$ ($H_1$) and
\\
\verb+       prove S!different(e1,e2) /\ belongs(e1,h::t)+ 
& let us prove $e_1 \neq e_2 \land e_1 \in h::t$. Two cases are possible\\
\verb+   <3>1 hypothesis C1 : S!equal(e2,h),+ & and we prove the property for these
two cases.  If $e_2=h$,  \\
\verb+        prove S!different(e1,e2) /\ belongs(e1,h::t)+ & \\
\verb+    <4>1 prove S!different(e1,e2)+& then we have $e_1 \neq e_2$ \\
\verb+         by definition of release+& since, by definition of $\ominus$,
$h::t \ominus e_2 = t \ominus e_2$, so by $H_1$,  \\
\verb+            hypothesis H1, C1, HI+& we get  $e_1 \in t \ominus e_2$ and by induction hypothesis $e_1 \neq e_2$.\\
\verb+    <4>2 prove belongs(e1,h::t)+& In remains to prove $e_1 \in
h::t$. \\
\verb+         <5>1 prove belongs(e1,t)+ &  Indeed, we have $e_1 \in
t$ since, by definition of $\ominus$, \\
\verb+              by hypothesis HI, C1, H1+ & 
$h::t \ominus e_2 = t \ominus e_2$, so by $H_1$, we get  $e_1 \in t \ominus e_2$
 \\
\verb+                 definition of release+ & 
and by induction hypothesis $e_1 \in t$.
\\
\verb+         <5>f qed by step <5>1+ & Hence, by definition of
\verb+from_list+ and by\\ 
\verb+                     property belongs_spec+ & property
\verb+belongs_spec+, \\
\verb+                     definition of from_list+ & we get $e_1 \in h::t$.\\
\verb+    <4>f conclude+ & Hence, when $e_2=h$, we have $e_1 \neq e_2
\land e_1 \in h::t$.\\
\verb+   <3>2 hypothesis C2 : not S!equal(e2,h),+ & Now, let us suppose
that $e_2 \neq h$. Two subcases are \\
\verb+        prove S!different(e1,e2) /\ belongs(e1,h::t)+ & possible
and we prove the property for both cases.\\
\verb+    <4>1 hypothesis C21 : S!equal(e1,h),+& If $e_1=h$,\\
\verb+         prove S!different(e1,e2) /\ belongs(e1,h::t)+ & \\
\verb+     <5>1 prove S!different(e1,e2)+ & then $e_1 \neq e_2$ \\
\verb+          by hypothesis C2, C21+ & since $e_2 \neq h$\\
\verb+             property S!equal_symmetric, S!equal_transitive,+ &
and by properties of equality.\\
\verb+                      S!same_is_not_different+ & \\
\verb+     <5>2 prove belongs(e1,h::t)+ & Furthermore we have $e_1 \in h::t$
(since $e_1=h$) \\
\verb+          by property belongs_spec, S!equal_symmetric+ & by
symmetry of equality and by definition\\
\verb+             hypothesis C21 definition of from_list+ & 
of
\verb+from_list+ and property
\verb+belongs_spec+.
\\
\verb+     <5>f conclude+ & \\
\verb+    <4>2 hypothesis C22 : not S!equal(e1,h),+ & Now, let us
suppose that $e_1 \neq h$.\\
\verb+         prove S!different(e1,e2) /\ belongs(e1,h::t)+ & \\
\verb+     <5>1 prove belongs(e1, h::release(t, e2))+ & We have $e_1
\in h::(t \ominus e_2)$ since $e_2 \neq h$\\
\verb+          by hypothesis H1, C2+ & 
and $e_1
\in h::t \ominus e_2$ ($H_1$) and by\\
\verb+             definition of release+ & definition of $\ominus$,
we get
$h::t \ominus e_2 = h::(t \ominus e_2)$.
\\
\verb+     <5>2 prove belongs(e1,release(t,e2))+ & Furthermore, it follows
 $e_1 \in t \ominus e_2$\\
\verb+          by step <5>1 hypothesis C22+ & since $e_1 \neq h$\\
\verb+             definition of belongs+ & and by definition of the
membership relation and \\
\verb+             property belongs_spec, S!equal_symmetric+ &
by property \verb+belong_spec+ and by symmetry of equality. \\
\verb+     <5>3 prove S!different(e1,e2) /\ belongs(e1,t)+ & 
Hence, by induction hypothesis, \\
\verb+          by step <5>2 hypothesis HI+ & we get $e_1 \neq e_2
\land e_1 \in t$.\\
\verb+     <5>4 prove belongs(e1,h::t)+ & 
From $e_1 \in t$, we get $e_1 \in h::t$
\\
\verb+          by step <5>3 definition of belongs+ &  by definition of the
membership relation \\
\verb+             property belongs_spec+ & and by property \verb+belong_spec+.\\
\verb+     <5>f conclude+ & \\
\verb+    <4>f conclude+ & Hence, when $e_2 \neq h$, we also have $e_1 \neq e_2
\land e_1 \in h::t$.\\
\verb+   <3>f conclude+ & \\
\verb+  <2>2 hypothesis H2: S!different(e1,e2) /\ belongs(e1,h::t),+
& ($\Leftarrow$) Let us suppose that $e_1 \neq e_2 \land e_1 \in
h::t$ ($H_2$) \\
\verb+       prove belongs(e1,release(h::t,e2))+ & and let us prove
that $e_1 \in h::t \ominus e_2$ \\
\verb+   <3>1 hypothesis C1 : S!equal(e1,h),+ & Two cases are
possible. If $e_1=h$,\\
\verb+        prove belongs(e1,release(h::t,e2))+ & \\
\verb+    <4>1 prove not S!equal(e2,h)+ & then $e_2 \neq h$\\
\verb+         by hypothesis H2, C1+ &  by
hypothesis $H_2$\\
\verb+            property S!equal_transitive, S!equal_symmetric,+ 
& and by properties of \\
\verb+                     S!same_is_not_different+
& equality.\\
\verb+    <4>2 prove release(h::t,e2)=h::release(t,e2)+ & 
Hence it follows $h::t \ominus e_2 = h:: (t \ominus e_2)$ \\
\verb+         by step <4>1 definition of release+ & by definition of $\ominus$.\\
\verb+    <4>3 prove belongs(e1,h::release(t,e2))+ & Furthermore, we
get $e_1 \in h::(t \ominus e_2)$ \\
\verb+         by definition of belongs hypothesis C1+ & since $e_1=h$
and by definition of the membership relation \\
\verb+            property  belongs_spec, S!equal_symmetric+ & 
and by property \verb+belongs_spec+ and symmetry of equality.\\
\verb+    <4>f qed by step <4>2, <4>3+ & Hence, when $e_1=h$, we have $e_1 \in h::t \ominus e_2$.\\
\verb+   <3>2 hypothesis C2 : not S!equal(e1,h),+ & Now, let us suppose
that $e_1 \neq h$.\\
\verb+        prove belongs(e1,release(h::t,e2))+ & \\
\verb+    <4>1 prove belongs(e1,t)+ & Then we get $e_1 \in t$ since
$e_1 \in h::t$ and $e_1 \neq h$ \\
\verb+         by definition of belongs hypothesis C2, H2+ & and by
definition of the membership relation, \\
\verb+            property belongs_spec, S!equal_symmetric+ & and by
property \verb+belongs_spec+ and symmetry of equality.\\     
\verb+    <4>2 prove belongs(e1,release(t,e2))+ & Hence, by induction
hypothesis, and since
$e_1 \neq e_2$ ($H_2$)\\
\verb+         by step <4>1 hypothesis H2, HI+ & it follows $e_1 \in t
\ominus e_2$\\
\verb+    <4>3 prove release(h::t,e2) = release(t,e2)+ & \\
\verb+               \/ release(h::t,e2) = h::release(t,e2)+ & 
Moreover, by definition of $\ominus$ we have\\
\verb+         by definition of release+ & 
$h::t \ominus e_2 = t \ominus e_2$ or $h::t \ominus e_2 = h::(t \ominus e_2)$.
\\
\verb+    <4>f qed by step <4>3, <4>2 + & Hence, when $e_1 \neq h$, we have $e_1 \in h::t \ominus e_2$\\
\verb+             property belongs_spec definition of from_list+ & 
(by definition of
\verb+from_list+ and by
property
\verb+belongs_spec+). \\
\verb+   <3>f conclude+ & \\
\verb+  <2>f conclude + & This concludes the inductive step.\\
\verb+ <1>f conclude + & This concludes the proof by induction. \\
 \verb+<0>f conclude;+ & This concludes the proof.\\
\hline
\end{tabular}
}
\end{center}
\caption{Proof of {\footnotesize \tt release\_spec}}\label{proof_release_spec}
\end{table}

Within FoCaLiZe, a collection can be built upon a completely defined species. This means
that every method must be defined. In other words, in a collection, every 
operation has an implementation, and every theorem is formally proved.
In addition, a collection is ``frozen'': it cannot be used as a parent
of a species in the inheritance graph. Moreover, to ensure modularity
and abstraction, the carrier of a collection is hidden: seen from the
outside, it becomes an abstract type. This means
that any software component dealing with a collection will only be
able to manipulate it through the operations it
provides (i.e. its methods). This point is especially important since
it prevents other
software components from breaking representation invariants required by the
internals of the collection.

\section{Conclusion}

Using a computer to teach discrete mathematics at the undergraduate
level is usally done by considering functionnal programming languages
allowing students to formally express computational contents of
mathematical concepts by programs and to informally reason on these
programs. In this paper, we go one step futher by showing that
abstract specifications and proofs can also be implemented at this
level whitout assuming some advanced theoretical background.
Indeed, while teaching experiences using F-IDE are mostly done at master
level, we believe that such approaches can also be adopted for
beginning students. We claim here that teaching how to develop
software with  F-IDE to
beginners is essential to ease and to promote their use in industry.

\begin{figure}
  \begin{center}
    \includegraphics[width=9cm]{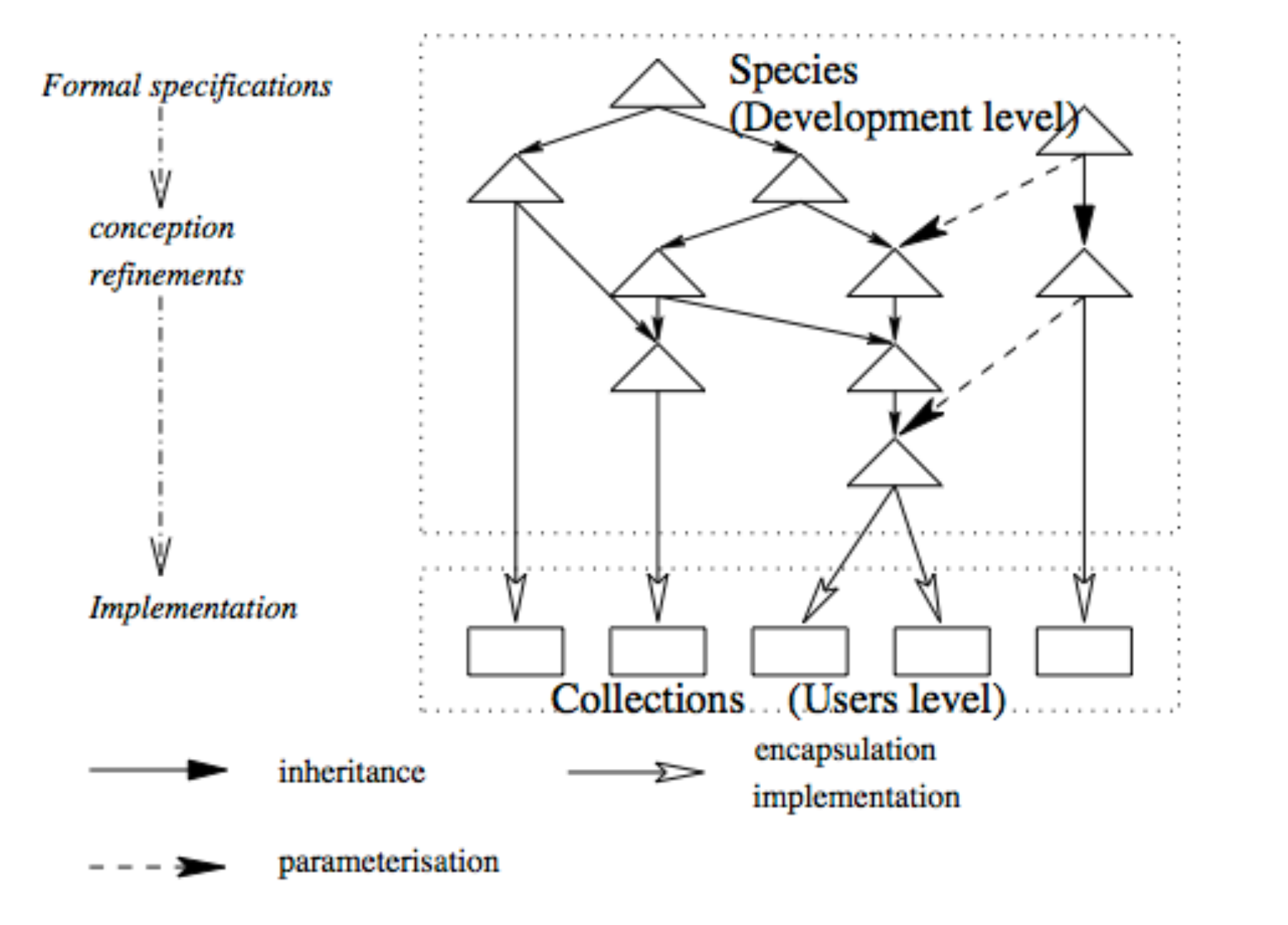}
    \caption{Formal development within \focalize}
    \label{fig:focal}
  \end{center}
\end{figure}

\focalize was conceived from the
beginning to help building systems with high
safety and security assurances. \focalize includes
a language based on firm theoretical
results~\cite{PrevostoJAR02}, with a clear semantics and provides an efficient
implementation -- {\it via} translation to OCaml.  It has
functional and object-oriented features and
provides means for the programmers to write formal proofs
of their
code
in a more or less detailed way
within a declarative proof language based on the Zenon automatic
theorem prover~\cite{conf/lpar/BonichonDD07}.
Zenon eases the task of writing formal proofs and
translates them into Coq for high-assurance checking.
\focalize also provides powerful features (such as
inheritance, parameterization and late-binding) that enable
a stepwise refinement methodology to go from specification all the way
down to executable code.
Indeed, thanks to
the main 
features of 
FoCaLiZe, a formal development can be organized as a
hierarchy (as illustrated in
figure~\ref{fig:focal}) which may have several roots:
the upper levels of the hierarchy are built during the specification
stage while the lower ones correspond to implementations.
Thus, \focalize unifies within the same language the formal modeling
work, the development of the code, and the certification proofs.
Very important is the ability in \focalize to have
specifications, implementations and proofs within the same
language, since it
eliminates the errors introduced between layers, at each switch
between languages, during the development cycle.
Other frameworks, like Atelier B~\cite{Abrial96a}, also aims at implementing
tools for making formal development a reality. \focalize doesn't follow
the same path, trying to keep the means of expression close to what
engineers usually know: a programming language.
Of course, nowadays, proof assistants also provide some features for
structuring code (module systems, type classes, etc), but most of them
still cannot be used
to obtain efficient programs. 
Compilation of \focalize developments leads to efficient OCaml programs
(which are not obtained by extracting computational contents of
proofs). It is this focus on efficiency that makes \focalize a real programming
language. To our knowledge, only the Agda~\cite{conf/tphol/BoveDN09}
programming language, based on dependent types and compiling {\it via}
Haskell, has a comparable mix of features.
Note that the \focalize language is also based on a
dependent type language, but with some restrictions on dependencies.
Furthermore, \focalize
provides several automatic tools to ease the generation of programs
from specifications, the generation of
documentation, and the production of
test suites~\cite{DBLP:conf/icsoft/CarlierDG10}.

For all these reasons, we think that \focalize is not only well suited
to develop critical systems but is also a good framework to teach both
computer science and discrete mathematics courses.
For example, we have already
used~\cite{Coq-Ens} \focalize (together with Coq) to teach (at a
master level) semantics of
object-oriented features of programming languages. 
In this paper, we consider \focalize as a teaching tool at the
undergraduate level and illustrate our approach with
a small development introducing very basic concepts 
of discrete mathematics and showing
how to mix both formal
methods and discrete mathematics courses.
Indeed,  \focalize provides
an environment simple enough to be 
usable by most students at university (even if they are not fully
acquainted with theoretical concepts such as 
higher-order logics), in particular by making development
of correct proofs as easy as possible and as readable  as possible.
Moreover, \focalize leads to stress the process of
abstraction through the construction, step by step, of problem
solutions from their specifications. This can be helpful
to improve the learning process of discrete mathematics but also to
show to students that
computer
science involves a lot of mathematical activities and {\it vice versa}.

In addition to pedagogical benefits, 
we believe that teaching how to use F-IDE as early as possible leads to
raise the level of mathematical rigor for computer science so as to
ensure that formal methods are perceived as valid professional disciplines by
students. 
Formal methods will be of
increasing value in computer and software engineering (especially for 
safety-critical, security-sensitive, and embedded systems)
and we think that education is one challenge to take up 
in order to promote the dissemination of formal methods in software industry.
\focalize includes a
computer algebra library, mostly developed by
R.~Rioboo~\cite{calc01,DBLP:journals/amai/Rioboo09}, which
implements mathematical structures up to multivariate 
polynomial rings and includes complex algorithms
with
performance comparable to the best computer algebra systems in
existence.
Hence, as future works, we believe that \focalize could be used 
to develop a complete discrete
mathematics course.


\paragraph{Acknowledgments}
The authors would like to thank Renaud Rioboo for his help  and
for enlightening discussions about how to program with FoCaLiZe
and about teaching discrete mathematics.

\bibliographystyle{eptcs}
\bibliography{fbibli}
\end{document}